\documentclass[aps,showpacs,amsmath,amssymb,floatfix,pre,twocolumn]{revtex4-1}
\usepackage{graphicx}
\usepackage{rotating}
\usepackage{longtable}
\usepackage{supertabular}
\usepackage{latexsym}
\usepackage{amssymb}
\usepackage{float}
\usepackage{dcolumn}
\usepackage{bm}

\begin{document}

\title{Non-universal behavior of helicity modulus in a dense defect system}
\author{Suman Sinha}
\email{suman.sinha@saha.ac.in}
\affiliation {Theoretical Condensed Matter Physics Division,\\
Saha Institute of Nuclear Physics, 1/AF Bidhannagar,\\
Kolkata - 700064, India}

\begin{abstract}
Extensive Monte Carlo simulation has been performed on a 2D modified XY model which behaves
like a dense defect system. Topological defects are shown to introduce disorders in the
system which makes the helicity modulus jump non-universal. The results corroborate the 
experimental observation of non-universal jump of the superconducting density in high-$T_c$
superconducting films.
\end{abstract}


\maketitle

The two-dimensional (2D) XY model of planar spins has been the subject of intense
interest since the last four decades. The 2D XY model was believed to be without 
a phase transition for a long time until Kosterlitz and Thouless (KT) proved that 
a phase transition indeed occurs and clarified its topological nature \cite{kt,kos}.
KT transition is a continuous phase transition
of infinite order from a high-temperature isotropic phase with exponential decay of 
spin-spin correlation functions to a low-temperature pseudo-long-range or quasi-long-
range order (QLRO) phase with power law decay of spin-spin correlations.  
KT pictured this transition in terms of vortex unbinding. In 
the low-temperature phase the charges (vortices) are bound together into dipole-pairs
while in the high-temperature phase some dipole-pairs are broken. KT
theory \cite{kos} leads to the famous universal jump prediction \cite{nelkos} of the 
helicity modulus \cite{ohta} or equivalently of the superfluid density \cite{nelkos,min1}.
Kosterlitz RG equations for the 2D Coulomb gas (CG) are constructed in the low-temperature phase and are
valid in the limit of small particle densities \cite{min2}. Later Minnhagen has suggested on
the basis of a new set of RG equations for the 2D CG that the conclusions based on Kosterlitz
RG equations may break down for larger dipole-pair fugacities \cite{min3,min4}. As a result
charge unbinding transition with non-universal jumps may, in principle, be possible and it
was shown that for higher particle densities, the charge-unbinding transition is first order
\cite{minwal}. Later Zhang $et. al$ \cite{zhang} on the basis of sine-Gordon (SG) field theory, 
showed that the nature of the transition in the dense 2D classical CG is of discontinuous first order
type. The evidence of a first order phase transition for higher particle densities was 
supported by Monte Carlo (MC) simulations \cite{cailev,leetei1,leetei2}. Such first order phase 
transitions are related to high-$T_c$ superconductivity \cite{korshu}. Leemann $et. al$, in
their experimental measurements of the inverse magnetic penetration depth $\Lambda^{-1}$ in thin
films of $\tt YBa_2Cu_3O_7$ \cite{leeman}, observed that the jump of the superconducting density (or 
equivalently the helicity modulus) did not obey the universal prediction of KT theory. These systems 
thus cannot be described by conventional XY model. Mila \cite{mila} attempted to understand the
the non-universal jump of the superconducting density in thin films of high-$T_c$ superconductors
in terms of XY model with a modified form of interaction potential. Such kind of model was first 
introduced by Domany and co-workers \cite{dss}. They introduced an extension of the 2D XY model
where the classical spins (of unit length), located at the sites of a square lattice and free 
to rotate in a plane, say the XY plane (having no Z-component) interact with nearest-neighbors
through a modified potential
\begin{equation}
V(\theta)=2J\Big[1-\Big(\tt cos^2{\frac{\theta}{2}}\Big)^{p^2}\Big]
\label{eqn1}
\end{equation}
where $\theta$ is the angle between the nearest neighbor spins, $J$ is the coupling constant 
and $p^2$ is a parameter used to
alter the shape of the potential, or in other words, $p^2$ controls the nonlinearity of the 
potential well, although variation in $p^2$ does not disturb the essential symmetry of the Hamiltonian.
For $p^2=1$, the potential reproduces the conventional XY model while for large values of $p^2$
(say $p^2=50$), the model behaves like a dense defect system \cite{ssskr3} and gives rise to a first
order phase transition as all the finite size scaling (FSS) rules for a first order phase transition 
were seen to be nicely obeyed \cite{ssskr2}. The first order phase transition is associated with a 
sharp jump in the average defect pair density \cite{him,ssskr3}. The change in the nature of the 
phase transition with the additional parameter $p^2$ is in contradiction with the prediction of RG
theory according to which systems in the same universal class (having same symmetry of the order
parameter and same lattice dimensionality) should exhibit the same type of phase transition with
identical values of critical exponents. In this context, I refer to the work of Curty and Beck
\cite{curty} who showed that in three dimension (3D), 
continuous phase transition can be preempted by a first order one.

Enter and Shlosman finally provided a rigorous
proof \cite{es,es2} of a first order phase transition in various SO($n$)-invariant $n$-vector models
which have a deep and narrow potential well. The model defined by Eqn. (\ref{eqn1}) is a member of
this general class of systems. Moreover, Enter and Shlosman argued that in spite of the order parameter
in the 2D systems with continuous energy spectrum being predicted to vanish by Mermin-Wagner theorem
\cite{merwag}, the first order transition is manifested by the long range order in higher-order
correlation functions. Recently Sinha and Roy \cite{ssskr2} verified this argument by numerical 
simulations and showed that while the lowest-order correlation function decays to zero, the next
higher-order correlation function has a finite plateau. Later they investigated the role of topological
defects on the phase transition exhibited by the model described by Eqn. (\ref{eqn1}) by means of
extensive MC simulations and 
observed that the system appears to remain ordered at all temperatures when configurations 
containing topological defects are not allowed to occur \cite{ssskr3}.

However, the connection of spin-stiffness (helicity modulus) with topological defects has not yet
been studied for systems exhibiting first order transition. The present Communication aims at studying this
aspect of phase transition which allows us to check whether the universal jump predicted in the
KT picture is valid in systems defined by Eqn. (\ref{eqn1}) too or not. The present paper also
explores the fact how disorder influences the properties of phase transition in these 2D systems.
The effect of disorder on the KT transition has become relevant since the experimental observation 
of superconductor-insulator transition in thin disordered films \cite{dis1,dis2}.

For the purpose of investigation, I choose the model defined by the interaction potential
given by Eqn. (\ref{eqn1}). I have found that the transition is associated with a non-universal 
drop in the helicity modulus. I also find that when the number density of topological defects 
increases rapidly, disorder is introduced into the system. The additional parameter $p^2$ 
plays the role of disorder here.

\begin{figure}[!h]
\begin{center}
\resizebox{80mm}{!}{\rotatebox{-90}{\includegraphics[scale=1.2]{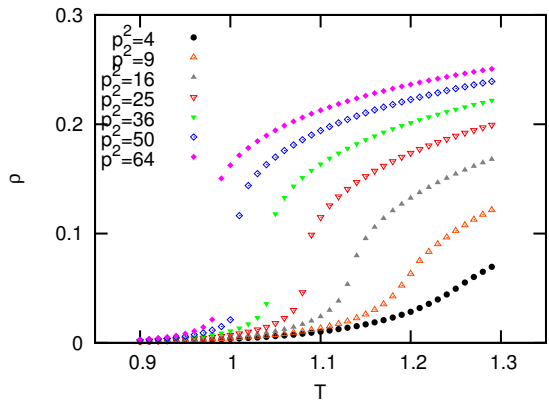}}}
\end{center}
\caption{(Color online) Average defect pair density $\rho$ plotted against dimensionless temperature $T$
for $L=64$ for various values of $p^2$}
\label{defdenvst}
\end{figure}

The variation of average defect pair density ($\rho$) with the dimensionless temperature $T$ 
for lattice size $L=64$ is shown in Fig. \ref{defdenvst}. The coupling constant $J$ (in Eqn.
(\ref{eqn1})) has been conventionally set to unity. 
The method for calculating average defect pair density and
the simulation techniques are discussed in Ref. \cite{ssskr3}. A sharp variation of $\rho$ as 
$T$ increases through the transition temperature $T_c(p^2)$ is observed. $\rho$
is found to show a sharp jump at $T_c(p^2)$, particularly for large values of $p^2$. 
So, for strong enough nonlinearity, there is a sudden proliferation in the average defect pair density
and the system under investigation behaves like a dense defect system for large values of $p^2$.

\begin{figure}[!h]
\begin{center}
\resizebox{80mm}{!}{\rotatebox{-90}{\includegraphics[scale=1.2]{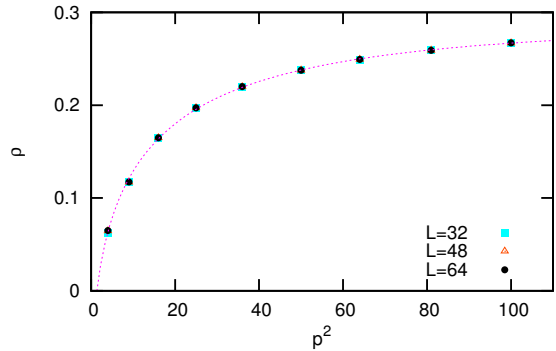}}}
\end{center}
\caption{(Color online) Average defect pair density $\rho$ plotted as a function of $p^2$ at $T=1.28$ for three system sizes.
 The best fit corresponds to $L=64$. The error bars are smaller than the dimension of the symbols used for
plotting.}
\label{p2vsdd}
\end{figure}

Next I plot the average defect pair density ($\rho$) as a function of the parameter $p^2$, shown
in Fig. \ref{p2vsdd}. The plot is for three different system sizes at a temperature $T=1.1200$, 
which is above the transition temperature of the model for $p^2=50$. The data for $\rho$ versus $p^2$ are
nicely fitted by the following expression
\begin{equation}
\rho(T)=\rho_{\tt max}-\alpha(T){\tt exp}(-\delta \sqrt p^2)
\label{eqn2}
\end{equation}
Eqn. (\ref{eqn2}) takes into account both vortices and anti-vortices. $\rho$ increases monotonically 
with $p^2$ and saturates to a maximum value $\rho_{\tt max} \approx 0.28$, which is independent of the 
temperature. The values of $\rho_{\tt max}$ for
the three system sizes are listed in Table \ref{table1}. There is no significant system size dependence of the
\begin{table}[!h]
\caption{parameters for the fit of $\rho(T)=\rho_{\tt max}-\alpha(T){\tt exp}(-\delta \sqrt p^2)$
for different $L$}
\begin{center}\
\begin{tabular}{|c|c|c|c|}
\hline
$L$ &$\rho_{\tt max}$ &$\alpha(T)$ &$\delta$ \\
\hline
$32$ &$0.2866\pm 0.002$ &$0.411\pm 0.005$ &$0.301\pm 0.008$ \\
\hline
$48$ &$0.2876\pm 0.002$ &$0.406\pm 0.006$ &$0.297\pm 0.009$ \\
\hline
$64$ &$0.2878\pm 0.002$ &$0.406\pm 0.006$ &$0.296\pm 0.009$ \\
\hline
\end{tabular}
\end{center}
\label{table1}
\end{table}
parameters as is evident from Table \ref{table1}. The maximum defect density ($\rho_{\tt max}$) is 
usually achieved in the high-temperature limit ($T \to \infty$) but here it is achieved in the
high-$p^2$ limit ($p^2 \to \infty$). Therefore it seems reasonable to interpret the parameter
$p^2$ to play the role of disorder. In the high-$p^2$ limit, the system contains only vortex 
excitations. This means that in the high-$p^2$ limit, the system must be disordered even at very
small temperature and consequently the transition temperature decreases with increase in $p^2$ which
we observe in Fig. \ref{defdenvst}. In other words, as we increase the nonlinearity $p^2$, 
 the influence of disorder becomes stronger and a tendency  
of a first order transition with a sudden proliferation of topological defects develops.

I now concentrate in the behavior of spin-stiffness (helicity modulus). Helicity
modulus, introduced by Fisher, Barber and Jasnow \cite{fbj},  
is a thermodynamic function which measures the ``rigidity'' of an isotropic system under
an imposed phase twist. The free energy difference between the twisted and the periodic
boundary conditions is proportional to the helicity modulus ($\gamma$). Thus the general
definition of helicity modulus may be regarded as \cite{cl}
\begin{equation}
\gamma=\displaystyle\lim_{L\to\infty}2L^{2-d}~\frac{F(\omega)-F(0)}{\omega^2}
\label{eqn3}
\end{equation}
where $\omega$ is the angle of twist and $d$ is the spatial dimension. In the present
case, the twisted boundary condition being anti-periodic and $d$ being $2$, the 
definition of $\gamma$ simplifies to 
\begin{equation}
\gamma=2~\frac{F(\pi)-F(0)}{\pi^2}
\label{eqn4}
\end{equation}
Anti-periodic boundary condition is imposed only along one direction, say along X-direction.
Our calculation of $\gamma$ involves a direct simulation of Eqn. (\ref{eqn4}) using
multiple reweighting histogram method, a sophisticated MC technique first proposed by
Ferrenberg and Swendsen \cite{fs,nb}. In the simulations, $10^7$ MC steps per site were
used for computing the raw histograms and $10^6$ MC steps per site were taken for 
equilibration.

\begin{figure}[!h]
\begin{center}
\resizebox{80mm}{!}{\rotatebox{-90}{\includegraphics[scale=1.2]{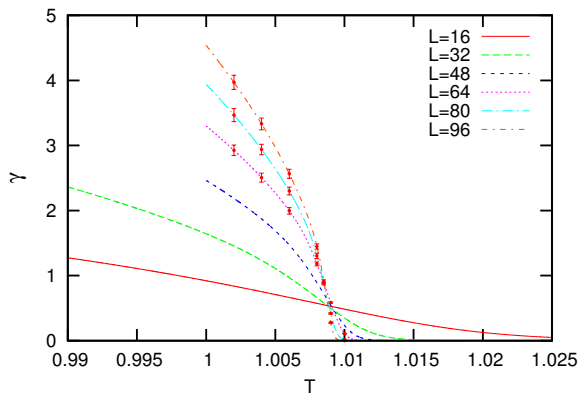}}}
\end{center}
\caption{(Color online) Helicity modulus $\gamma$ against temperature ($T$) for various lattice sizes
for $p^2=50$ with the errorbars shown (for three lattice sizes).}
\label{helivst}
\end{figure}

The variation of $\gamma$ with temperature for various lattice sizes is displayed in
Fig. \ref{helivst}. The transition is signaled by an abrupt decrease of the helicity
modulus in the vicinity of the transition temperature as the temperature is increased and
the drop at the transition gets steeper as the system size(L) is increased. It is also 
manifested that instead of the subtle and the smooth KT transition, the
transition coincides with a non-universal jump in $\gamma$. The physical picture can be
explained as follows. As $p^2$ increases, short-scale fluctuations (rotation of separate spins 
by large angles) are favored over long-wavelength fluctuations (spin waves and vortices) and 
when this happens disordering is induced at a lower temperature than the KT transition 
temperature, but the vortex-vortex interaction still
remains  stronger at that lower temperature, thus making the helicity modulus 
jump non-universal. The ratio of $\gamma /T$ at $T=T_c$ for different $L$
is listed in Table \ref{table2}. I point out that Minnhagen \cite{min3} also showed the
possibility of a KT transition in a 2D CG with a non-universal jump in $\gamma$.
\begin{table}[!h]
\caption{$\gamma/T$ values at $T=T_c$ for different L} 
\begin{center}\
\begin{tabular}{|c|c|c|c|c|c|c|}
\hline
$L$ &$16$ &$32$ &$48$ &$64$ &$80$ &$96$ \\
\hline
$\gamma /T$ &$0.5742$ &$0.6049$ &$0.6569$ 
&$0.7192$ &$0.6971$ &$0.6454$ \\
\hline
\end{tabular}
\end{center}
\label{table2}
\end{table}

\begin{figure}[!h]
\begin{center}
\resizebox{80mm}{!}{\rotatebox{-90}{\includegraphics[scale=1.2]{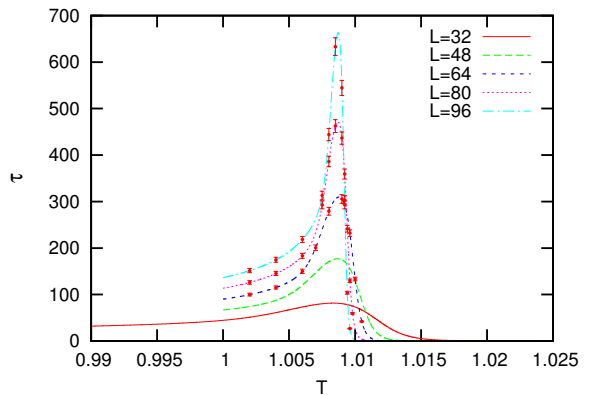}}}
\end{center}
\caption{(Color online) Derivative of $\frac{1}{2} \beta \gamma(\beta)$ as a function of temperature $T$ for different
lattice sizes for $p^2=50$ with the errorbars shown (for three lattice sizes).}
\label{derihelivst}
\end{figure}

I have also computed the temperature derivative ($\tau$) of the helicity modulus. The 
internal energy difference under anti-periodic and periodic boundary conditions gives
the derivative of helicity modulus in the form
\begin{equation}
\tau=\frac{1}{2}~ \frac{d}{d\beta}~[\beta \gamma(\beta)]=\frac{{\langle}E{\rangle}_a-{\langle}E{\rangle}_p}{\pi^2}
\label{eqn5}
\end{equation}
The MC data for the right hand size of Eqn. (\ref{eqn5}) as a function of temperature for
different lattice sizes are shown in Fig. \ref{derihelivst}. The transition is manifested by a 
huge peak height in $\tau$ and the data display a divergent behavior with increasing $L$, 
indicative of a discontinuous jump in $\gamma$ in an infinite lattice.

\begin{figure}[!h]
\begin{center}
\resizebox{80mm}{!}{\rotatebox{-90}{\includegraphics[scale=1.2]{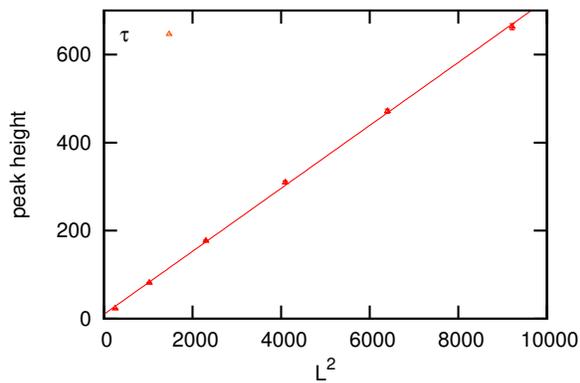}}}
\end{center}
\caption{(Color online) The peak heights of $\tau$ plotted against $L^2$ with the linear fit
represented by the straight line. The error bars for most points are smaller than
the dimension of the symbols used for plotting.}
\label{fsstau}
\end{figure}

Now I present the FSS of $\tau$. Since $\tau$ is a response function like specific heat
($C_v$) or susceptibility ($\chi$), it is expected to show an identical behavior in scaling
as for $C_v$ or $\chi$. From Fig. \ref{fsstau}, where the maxima of $\tau$ are plotted 
against $L^2$, it is clear that the peak heights of $\tau$ scale as $L^d$ which confirms
the first order nature of the present model for $p^2=50$. We recall here that for first
order transition standard scaling rule for $C_v$ goes like $C_v \sim L^d$ \cite{binder}.

In this Communication I have used extensive MC simulation to show that for strong enough
nonlinearity (i.e., for large values of $p^2$) in the interaction potential of
Eqn. (\ref{eqn1}), there is a sudden proliferation of topological defects which makes
the system disordered. Consequently the transition is associated with a discontinuous 
non-universal jump in the helicity modulus. Thus our simulation has given some support
to the idea that the type of phase transition in thin superconducting films may be changed 
due to influence of disorder. As high-$T_c$ superconducting films are believed to have a 
irregular structure, it seems reasonable to relate the non-universal jump to disorder.

However, some studies \cite{mila,knops,him1}, mostly based on RG 
analysis of Migdal-Kadanoff type contested the first order nature of the transition in the
model defined by Eqn. (\ref{eqn1}). Since renormalization arguments hold good only for
small disorder, the possibility that disorder may change the nature of phase transition 
always remains there. Perhaps this is the reason behind the different interpretation of
results by the authors of Ref. \cite{mila,knops,him1}.

Finally, the present work could shed light on the nature of 2D melting which remains
controversial for decades. Experimental works and numerical simulations favor a KT-like
transition in some cases and a discontinuous one in others \cite{stra}. After all, it is
possible that the nature of the melting transition in 2D depends on the specific system
and the parameters of the model which in turn translate into different values of the 
nonlinearity parameter $p^2$.

I end this paper with a comment. It is observed in Fig. \ref{helivst} that the graphs 
for helicity modulus ($\gamma$) against temperature for different lattice sizes 
intersect at a point which is the transition temperature of the model (within an
error of $0.03\%$). I offer no explanation for this interesting result and
this issue is left for future research.

The author is grateful to S. K. Roy for many fruitful discussions and a number of
suggestions after critically reading this manuscript.

\end{document}